\documentstyle[aps,prb]{revtex}
\begin{document}

\twocolumn[\hsize\textwidth\columnwidth\hsize\csname
@twocolumnfalse\endcsname

\title{Jellium surface energy beyond the local-density approximation:
Self-consistent-field calculations}
\author{J. M. Pitarke$^1$ and A. G. Eguiluz$^2$} 
\address{$^1$Materia Kondentsatuaren Fisika Saila, Zientzi
Fakultatea, Euskal Herriko  Unibertsitatea,\\ 644 Posta kutxatila, 48080 Bilbo,
Basque Country, Spain\\
and Donostia International Physics Center (DIPC) and Centro Mixto
CSIC-UPV/EHU,\\
Donostia, Basque Country, Spain\\
$^2$Department of Physics and Astronomy, The University of Tennessee, Knoxville,
TN 37996-1200\\
and Solid State Division, Oak Ridge National Laboratory, Oak Ridge,TN
37831-6032}

\date\today
\maketitle

\begin{abstract}
We report extensive self-consistent calculations of jellium surface energies,
by going beyond the local-density approximation. The density-response
function of a bounded free-electron gas is evaluated within the random-phase
approximation, with use of self-consistent electron density profiles. The
exchange-correlation energy is then obtained from an exact adiabatic
fluctuation-dissipation formula. We also investigate quantum-size effects and
the extrapolation of finite-slab calculations to the infinite-width limit.
\end{abstract}

]

\section{Introduction}

Density-functional theory (DFT),\cite{Hohenberg} in the local-density
approximation (LDA) and its various gradient-corrected forms,\cite{Sham} has been
the most widely used method to investigate the properties of solid
surfaces.\cite{Lang0,Lang} However, the rapidly varying electron density near the
surface does not justify {\it a priori} the use of local or quasilocal
approximations. In particular, these approximations are well known to yield an
inaccurate description of the position-dependent exchange-correlation (xc) hole
density, they lead to a surface barrier which does not present the correct
image-like asymptotic behavior, and there has been a long-standing controversy
about the extent to which the LDA accounts for the exact xc surface
energy.\cite{Harris,Lang75,Rasolt75,Wikborg,Langreth75,Perdews,Perdew79}
Recently, the wave-function-based Fermi
hypernetted-chain (FHNC)\cite{Kohn86} and quantum Monte
Carlo (QMC)\cite{Ceperley92,Acioli} predictions have been found to disagree with
modern density-functional calculations of jellium surface energies that go
beyond a local-density approximation.\cite{Zhang,Perdew1,Perdew2,Perdew3}

In this paper, we present extensive self-consistent calculations of jellium
surface energies. The dynamical density-response function of a
bounded free electron gas (FEG) is evaluated within the
random-phase approximation (RPA),\cite{Mahan} from the knowledge of the
non-interacting density-response function. The xc energy of jellium slabs is
then obtained via a coupling-constant integration over the interacting
density-response function. For comparison, we first consider a simple
non-selfconsistent  microscopic model of the surface, the so-called
infinite-barrier model (IBM), for which analytical insight is possible by virtue
of the simple nature of the one-electron wave functions. Full self-consistent
calculations are performed by employing either Kohn-Sham or Hartree orbitals to
construct the non-interacting density-response function. LDA calculations of
jellium surface energies are also carried out, within the same density-response
framework, and a comparison between our {\it local} and {\it non-local}
calculations indicate that the error introduced by the LDA is small.

Quantum-size effects (QSE)\cite{Schulte,Feibelman} on the surface energy of
jellium slabs are also investigated, and the extrapolation of finite-slab
calculations to the infinite-width limit is discussed. This issue is important,
in view of the potential application of this procedure to the case of more
complex slab calculations involving the use of the band structure and/or QMC
simulations.

\section{The jellium surface energy}  

We consider a jellium slab of thickness $a$ normal to the $z$ axis, consisting of
a fixed uniform positive background of density
\begin{equation}
n_+(z)=\cases{\bar n,&$-a/2\leq z\leq a/2$\cr\cr 0,& elsewhere,}
\end{equation}
plus a neutralizing cloud of interacting electrons.

The surface energy is defined as the energy per unit area required to split the
solid into two separate halves along a plane. For a solid that is translationally
invariant in the plane of the surface, DFT shows that\cite{Lang}
\begin{equation}
\sigma=\sigma_s+\sigma_{es}+\sigma_{xc},
\end{equation}
where $\sigma_s$ represents the kinetic surface energy of non-interacting
Kohn-Sham electrons
\begin{eqnarray}\label{eqs}
\sigma_s=&&{m\over 4\pi\hbar^2}\,\sum_{l=1}^{l_M}(E_F^2-\varepsilon_l^2)
-{\hbar^2\over 2m}\,{q_F^5\,a\over 10\pi^2}\cr\cr &&-\int_0^\infty
dz\,n(z)\left\{v_{eff}[n](z)-v_{eff}[n](0)\right\},
\end{eqnarray}
$\sigma_{es}$ is the electrostatic surface energy of all positive and
negative charge distributions
\begin{equation}\label{eqes}
\sigma_{es}={1\over 2}\int_0^\infty dz\,[n(z)-n_+(z)]\,\varphi(z),
\end{equation}
and $\sigma_{xc}$ is the xc surface energy
\begin{equation}\label{eqxc}
\sigma_{xc}=\int_0^\infty
dz\,n(z)\,\left\{\varepsilon_{xc}[n](z)-\varepsilon_{xc}^{unif}(\bar n)\right\}.
\end{equation}
Here, $n(z)$ is the electron density
\begin{equation}\label{eqden}
n(z)={m\over\pi\hbar^2}\,\sum_{l=1}^{l_M}(E_F-\varepsilon_l)\,|\phi_l(z)|^2,
\end{equation}
$v_{eff}[n](z)$ is the Kohn-Sham effective one-electron potential
\begin{equation}\label{effective}
v_{eff}[n](z)=\varphi(z)+v_{xc}[n](z),
\end{equation}
$\varphi(z)$ and $v_{xc}[n](z)$ being the electrostatic and xc potential,
respectively, $\varepsilon_{xc}[n](z)$ represents an xc energy per
particle at point $z$, and $\varepsilon_{xc}^{unif}(\bar n)$ is the xc energy per
particle of a uniform electron gas of density $\bar n$. $\varepsilon_l$ and
$\phi_l(z)$ are eigenvalues and eigenfunctions of the one-dimensional Kohn-Sham
hamiltonian describing motion normal to the surface, $l_M$ denotes the highest
occupied level, $E_F$ represents the Fermi level, and $q_F=(3\pi^2\bar
n)^{1/3}$ is the Fermi momentum. In the LDA, the xc surface energy is obtained by
simply replacing $\varepsilon_{xc}[n](z)$ by the xc energy
per particle of a uniform electron gas of density $n(z)$, i.e.,
\begin{equation}\label{eqxclda}
\sigma_{xc}^{LDA}=\int_0^\infty
dz\,n(z)\,\left\{\varepsilon_{xc}^{unif}[n(z)]-\varepsilon_{xc}^{unif}(\bar
n)\right\}.
\end{equation}
 
The {\it exact} xc energy per particle at point ${\bf r}$ can be obtained as
the energy due to the interaction between an electron at ${\bf r}$ and its
averaged xc hole:
\begin{equation}\label{exc}
\varepsilon_{xc}[n]({\bf r})={e^2\over 2}\int d{\bf r}'\,{\bar n_{xc}({\bf
r},{\bf r}')\over|{\bf r}-{\bf r}'|},
\end{equation}
where $\bar n_{xc}({\bf r},{\bf r}')$ is defined by
adiabatically  switching on the electron-electron interaction via a
coupling constant $\lambda$, i.e., $v^\lambda({\bf r}-{\bf r}')=\lambda e^2/|{\bf
r}-{\bf r}'|$ and by adding, at the same time, an external potential so as to
maintain the true  ($\lambda=1$) ground-state density in the presence of the
modified electron-electron interaction. One writes\cite{Langreth75}
\begin{equation}\label{nxc1}
\bar n_{xc}({\bf r},{\bf r}')=\int_0^1 d\lambda\,n_{xc}^\lambda({\bf r},{\bf
r}'),
\end{equation}
where $n_{xc}^\lambda({\bf r},{\bf r}')$ is the xc-hole density of a
fictitious  system at coupling strength $\lambda$. Using the
fluctuation-dissipation theorem,\cite{fd} one may write
\begin{eqnarray}\label{nxc2}
n_{xc}^\lambda({\bf r},{\bf r}')={1\over n({\bf
r})}[&&-{\hbar\over\pi}\int_0^\infty d\omega
\chi^\lambda({\bf r},{\bf r}';i\,\omega)\cr\cr
&&\left.-n({\bf r})\,\delta({\bf r}-{\bf r}')\right],
\end{eqnarray}
where $\chi^\lambda({\bf r},{\bf r}';\omega)$ represents the so-called
density-response function of the electron system.

Time-dependent DFT (TDDFT) shows that the {\it exact} density-response
function satisfies the integral equation\cite{DFT}
\begin{eqnarray}\label{chi}
\chi^\lambda&&({\bf r},{\bf r}';\omega)=
\chi^0({\bf r},{\bf r}';\omega)+\int d{\bf r}_1\int{\rm
d}{\bf r}_2\cr\cr
&&\times\chi^0({\bf r},{\bf
r}_1;\omega)\left\{v^\lambda({\bf r}_1,{\bf r}_2)
+f_{xc}^\lambda[n]({\bf r}_1,{\bf r}_2;\omega)\right\}\cr\cr
&&\times\chi^\lambda({\bf r}_2,{\bf r}';\omega),
\end{eqnarray}
where
\begin{equation}\label{kernel}
f_{xc}^\lambda[n]({\bf r},{\bf r}';\omega)={\delta v_{xc}^\lambda[n]({\bf
r},\omega)\over\delta n({\bf r}',\omega)},
\end{equation}
$v_{xc}^\lambda[n]({\bf r},\omega)$ being the exact frequency-dependent xc
potential,\cite{Runge} and $\chi^0({\bf r},{\bf r}';\omega)$, the
density-response function of non-interacting Kohn-Sham electrons, i.e.,
independent electrons moving in the effective potential
$v_{eff}[n]({\bf r})$ entering the Kohn-Sham equation of DFT.

If the interacting density-response function $\chi^\lambda({\bf r},{\bf
r}';\omega)$ is replaced by $\chi^0({\bf r},{\bf r}';\omega)$, Eq.
(\ref{exc}) yields the {\it exact} exchange energy per particle at point ${\bf
r}$, $\varepsilon_x[n]({\bf r})$. The RPA $\varepsilon_{xc}[n]({\bf r})$ is
derived by simply setting the xc kernel of Eq. (\ref{kernel}) equal to
zero.\cite{noterpa} In the so-called time-dependent local-density approximation
(TDLDA)\cite{tdlda} or, equivalently, adiabatic local-density approximation
(ALDA), the exact long-wavelength limit of the static ($\omega=0$) xc kernel is
used, i.e.,
\begin{equation}\label{kernellda}
f_{xc}^{\lambda,ALDA}[n]({\bf r},{\bf r}';\omega)={d
v_{xc}^{\lambda,unif}[n({\bf r})]\over dn({\bf
r})}\delta({\bf r}-{\bf r}'),
\end{equation}
$v_{xc}^{\lambda,unif}[n({\bf r})]$ being the xc potential of a uniform
electron gas of density $n({\bf r})$. 

For a uniform electron gas of density $n$, introduction of Eq. (\ref{nxc2})
into Eq. (\ref{nxc1}) and then Eq. (\ref{nxc1}) into Eq. (\ref{exc}) yields
\begin{eqnarray}\label{uniform}
\varepsilon_{xc}^{unif}(n)=&&{1\over 2}\int{d{\bf
q}\over(2\pi)^3}\,v(q)\cr\cr
&&\times\left[-{\hbar\over\pi\,n}\int_0^1d\lambda\int_0^\infty
d\omega\,\chi^\lambda(q,i\,\omega)-1\right],
\end{eqnarray}
where $v(q)$ and $\chi^\lambda(q,\omega)$ represent three-dimensional Fourier
transforms of the Coulomb potential and the density-response function,
respectively. If the interacting density-response function
$\chi^\lambda(q,\omega)$ is replaced by the Lindhard
function $\chi^0(q,\omega)$,\cite{Lindhard} Eq. (\ref{uniform}) yields the
{\it exact} exchange energy per particle of a uniform electron gas,
\begin{equation}
\varepsilon_x^{unif}(n)=-{3e^2\over
4\pi}\left(3\pi^2n\right)^{1/3},
\end{equation}
and the LDA exchange surface energy
\begin{equation}\label{xlda}
\sigma_x^{LDA}={3e^2\over 8\pi}\,\bar
n\,\int_0^\infty dz'\,\rho(z')\left[1-\rho^{1/3}(z')\right],
\end{equation}
where $\rho(z)=n(z)/\bar n$ represents the normalized electron density, and
$z'=2\,q_F\,z$. 

For a jellium slab, which is translationally invariant only in a
plane perpendicular to the $z$ axis, one finds
\begin{eqnarray}\label{eq1}
\varepsilon_{xc}&&[n](z)={1\over 2}\int{d{\bf q}_\parallel\over(2\pi)^2}\int
dz'\,v(z,z';q_\parallel)
\left[-{\hbar\over\pi\,n(z)}\right.\cr\cr
&&\left.\times\int_0^1d\lambda\int_0^\infty
d\omega\,\chi^\lambda(z,z';q_\parallel,i\,\omega)-\delta(z-z')\right],
\end{eqnarray}
where $v(z,z';q_\parallel)$ and $\chi^\lambda(z,z';q_\parallel,\omega)$
represent two-dimensional Fourier transforms of the Coulomb potential
and the density-response function, and ${\bf q}_\parallel$ is a wave vector
parallel to the surface.

To find $\chi^\lambda(z,z';q_\parallel,\omega)$, we
follow the method described in Ref.\onlinecite{Eguiluz}. We first assume that
$n(z)$ vanishes at a distance $z_0$ from either jellium edge,\cite{note1} and
expand the wave functions $\phi_l(z)$ in a Fourier sine series. We then
introduce a double-cosine Fourier representation for the density-response
function, which allows us to transform Eq. (\ref{chi}) into a matrix
equation that is solved numerically. The integrals of Eqs. (\ref{eqxc}) and
(\ref{eq1}) over the coordinate normal to the surface can be performed
analytically, and we find an explicit expression for the xc surface energy, in
terms of the eigenvalues $\varepsilon_l$ and the Fourier coefficients of the
eigenfunctions $\phi_l(z)$ (see Appendix A). 

\section{The infinite barrier model}

In this section we consider the simplest possible microscopic model of the
jellium slab, namely the so-called infinite barrier model (IBM).\cite{Bardeen}
This non-selfconsistent model has been widely
discussed,\cite{News,Griffin,Inglesfield} because of the simplicity of the
one-electron wave functions. Nevertheless, we are not aware of any theoretical
description of the extrapolation, within this model, of finite-slab
calculations to the infinite-width limit. Since this issue may be important, in
view of its potential application to the case of more complex self-consistent
calculations, we present below a detailed analysis of QSE on the surface energy
of an electron system confined in an infinite well.

In the IBM, the Kohn-Sham effective one-electron potential $v_{eff}[n](z)$ is
replaced by infinitely high potential walls at a distance $z_0$ outside the
jellium slab, 
\begin{equation}\label{eff}
v_{eff}(z)=\cases{0,& $-a/2-z_0\leq a/2+z_0$\cr\cr
\infty,& elsewhere,}
\end{equation}
with $z_0$ chosen so as to ensure charge neutrality, i.e.,
\begin{equation}\label{eqn0}
\rho_0\,(a+2z_0)\equiv 2\int_0^{a/2+z_0}dz\,\rho(z)=a,
\end{equation}
$\rho_0$ representing the normalized electron density averaged over the region
between the two infinite barriers.

Hence, the normalized Kohn-Sham one-electron wave functions and energies are
\begin{equation}
\phi_l(z)=\sqrt{2\over d}\,\sin\left({l\pi\over d}z\right)
\end{equation}
and
\begin{equation}\label{levels}
\varepsilon_l={\hbar^2\over 2m}\,\left({l\pi\over d}\right)^2,
\end{equation}
with
$d=a+2z_0$. Introduction of these eigenfunctions and eigenvalues into Eq.
(\ref{eqden}) yields the IBM electron density, and no attempt is then made to
alter $v_{eff}[n](z)$ using this computed density.

In the infinite-width (iw) limit and for $z>0$, Eq. (\ref{eqden})
yields the well-known result\cite{Lang}
\begin{equation}\label{rhoiw}
\left[\rho(z)\right]_{iw}=\left[1+3\,\tilde z^{-3}(\tilde  z\,\cos\tilde
z-\sin\tilde z)\right]\Theta(-\tilde z),
\end{equation}
where $\tilde z=2q_F(z-d/2)$. By carrying out the
integration of Eq. (\ref{eqn0}) one then easily finds
\begin{equation}\label{limit}
\left[\rho_0\right]_{iw}=1-{3\over 4x}
\end{equation}
and
\begin{equation}\label{zlim}
\left[z_0\right]_{iw}=(3/16)\lambda_F,
\end{equation}
where the dimensionless parameter $x=2d/\lambda_F$ has been introduced,
$\lambda_F=2\pi/q_F$ being the Fermi wavelength.

Instead, for a finite slab one finds
\begin{equation}\label{n0}
\rho_0={3\over 2x}\sum_{l=1}^{l_M}\left[1-\left({l\over
x}\right)^2\right]
\end{equation}
and
\begin{equation}\label{z0}
z_0=\lambda_F\,{x\over 4}\left\{1-{3\over
2x}\sum_{l=1}^{l_M}\left[1-\left({l\over x}\right)^2\right]\right\},
\end{equation}
with the highest occupied level $l_M$ being the largest integer less than or
equal to $x$.

The dependence of the average electron density $\rho_0$ on the parameter $x$, as
obtained from Eq. (\ref{n0}), is shown in Fig. 1. This figure clearly shows the
oscillatory character of this quantity, which is a consequence of the so-called
QSE first reported by Schulte.\cite{Schulte} As $d$ increases,
new subbands for the $z$ motion become occupied. One easily sees that such a
new subband falls below the Fermi level every time $d$ is a multiple of
$\lambda_F/2$, i.e., every time
$x$ is integer. When a new subband is pulled below the Fermi level, the parallel
Fermi sea built upon the newly occupied subband acquires more electrons, thereby
increasing $\rho_0$. However, this effect is eventually overcome by the fact that
all the subbands for the $z$ motion get deeper with increasing film thickness.
When $x$ has increased by unity, a new subband begins to be filled and a new
oscillation begins.

We define the threshold value $\left[\rho_0\right]_1$ of the normalized average
electron density $\rho_0$ corresponding to a jellium slab with $x$
integer. By using the Euler-Maclaurin summation
formula,\cite{abra} Eq. (\ref{n0}) with $x=n$ yields a local minimum,
\begin{equation}\label{xn}
\left[\rho_0\right]_1=1-{3\over 4x}\left(1+{1\over 3x}\right).
\end{equation}
We also define $\left[\rho_0\right]_2$ corresponding to a jellium slab with
$x=n+1/2$, and find
\begin{equation}\label{xnplus}
\left[\rho_0\right]_2=1-{3\over 4x}\left(1-{1\over 6x}\right).
\end{equation}
As $x$ increases, both Eqs. (\ref{xn}) and (\ref{xnplus}) approach the
infinite-width asymptotic behavior of Eq. (\ref{limit}). Although
$\left[\rho_0\right]_1$ and $\left[\rho_0\right]_2$  represent the actual value
of $\rho_0$ only at $x=n$ and $x=n+1/2$, respectively, they are exhibited in Fig.
1 for all values of $x$. This figure shows that Eqs. (\ref{xn}) and
(\ref{xnplus}) represent enveloping lines of the actual oscillating behavior of
$\rho_0$. Hence, the amplitude of the oscillations is
\begin{equation}
\left[\rho_0\right]_2-\left[\rho_0\right]_1={1\over 2x},
\end{equation}
thus decaying linearly with the thickness of the slab. Moreover, the
infinite-width limit of Eq. (\ref{limit}) is obtained from the enveloping lines
of Eqs. (\ref{xn}) and (\ref{xnplus}), as follows
\begin{equation}\label{average}
\left[\rho_0\right]_{iw}={\left[\rho_0\right]_1+2\left[\rho_0\right]_2\over 3}.
\end{equation}

Fig. 2 shows that a similar oscillatory behavior is exhibited by
the normalized electron density $\rho(z)$ at the center of the slab ($z=0$):
\begin{equation}\label{rho0}
\rho(0)={3\over x}\sum_{l=1}^{l'_M}\left[1-\left({2l-1\over x}\right)^2\right],
\end{equation}
where $l'_M$ is the largest integer less than or equal to $(x+1)/2$. Also
shown in Fig. 2 are the enveloping lines
\begin{equation}\label{rho1}
\left[\rho(0)\right]_1=1-{1\over x^2}
\end{equation}
and
\begin{equation}\label{rho2}
\left[\rho(0)\right]_2=1+{1\over 2 x^2},
\end{equation}
which represent the actual value of $\rho(0)$ for $x$ odd and even,
respectively. The infinite-width limit $\left[\rho(0)\right]_{iw}=1$ is
obtained as in Eq. (\ref{average}), with $\rho(0)$ instead of $\rho_0$.
 
At the threshold width for which the $n$th subband for the $z$ motion is first
occupied ($x=n$), the normalized average electron density $\rho_0$ is smaller
than in the infinite-width limit, i.e.,
$\left[\rho_0\right]_1<\left[\rho_0\right]_{iw}$, which requires that the
positive background be cut-off further from the infinite wells, i.e.,
$z_0$ is a local maximum. This is illustrated in Fig. 3, where the dependence of
the ratio
$z_0/\left[z_0\right]_{iw}$ on the parameter $x$ is exhibited, as obtained from
Eq. (\ref{z0}).  Also plotted in this figure are the enveloping lines
\begin{equation}\label{z1}
\left[z_0\right]_1=\left[z_0\right]_{iw}\left[1+{1\over 3x}\right]
\end{equation}
and
\begin{equation}\label{z2}
\left[z_0\right]_2=\left[z_0\right]_{iw}\left[1-{1\over 6x}\right],
\end{equation}
which coincide with the actual value of $z_0$ for $x$ integer ($x=n$) and for
$x=n+1/2$, respectively. An inspection of Eqs. (\ref{z1}) and
(\ref{z2}) easily shows that the infinite-width limit of Eq. (\ref{zlim}) is
obtained as in Eq. (\ref{average}), with $\rho_0$ replaced by $z_0$.

\subsection{Kinetic surface energy}

An explicit expression for the surface-energy term $\sigma_s$ is
obtained by substituting Eqs. (\ref{eff}) and (\ref{levels}) into Eq.
(\ref{eqs}). One finds
\begin{eqnarray}\label{sigmas}
\sigma_s=\left[\sigma_s\right]_{iw}&&\left\{10\,\sum_{l=1}^{l_M}
\left[1-\left({l\over x}\right)^2\right]\right.\cr\cr\cr
&&\left.-12\,\sum_{l=1}^{l_M}\left[1-\left({l\over x}\right)^4\right]\right\},
\end{eqnarray}
where
\begin{equation}\label{limits}
\left[\sigma_s\right]_{iw}={\hbar^2\over 2m}\,{q_F^4\over 80\pi}
\end{equation}
is the often-derived value of the IBM surface-energy term $\sigma_s$ of a
semi-infinite medium.\cite{Lang} For $x$ integer, the Euler-Maclaurin
summation formula now yields
\begin{equation}\label{env1}
\left[\sigma_s\right]_1=\left[\sigma_s\right]_{iw}\left[1-{1\over
3x}\left(4-{1\over x^2}\right)\right],
\end{equation}
while for  $x=n+1/2$ one finds
\begin{equation}\label{env2}
\left[\sigma_s\right]_2=\left[\sigma_s\right]_{iw}\left[1+{1\over
6x}\left(4-{1\over x^2}\right)-{1\over 8x^3}\right].
\end{equation}

The dependence of the ratio $\sigma_s/\left[\sigma_s\right]_{iw}$ on the
parameter $x$, as obtained from Eq. (\ref{sigmas}), is exhibited in Fig. 4.
Also plotted in this figure are $\left[\sigma_s\right]_1$ and
$\left[\sigma_s\right]_2$. While $\left[\sigma_s\right]_1$ represents the
actual surface-energy term $\sigma_s$ at the local minima ($x=n$), local
maxima slightly deviate from $x=n+1/2$, especially for the smallest values of
$x$. For large values of $x$, Eq. (\ref{env2}) approaches the upper enveloping
line of the true oscillating behavior of $\sigma_s$. Furthermore, one finds
\begin{equation}\label{sigmalim}
\left[\sigma_s\right]_{iw}={\left[\sigma_s\right]_1+2\left[\sigma_s\right]_2\over
3}+O(x^{-3}),
\end{equation}
where the first term quickly approaches the actual infinite-width limit of
Eq. (\ref{limits}), as shown in Fig. 4.

\subsection{Exchange surface energy}

The LDA exchange surface energy is obtained by simply introducing the
normalized electron density into Eq. (\ref{xlda}). In the infinite-width
limit $\rho(z)$ is that of Eq. (\ref{rhoiw}), and numerical integration
leads to the following result\cite{noteHarris}
\begin{equation}\label{xlimit1}
\left[\sigma_x^{LDA}\right]_{iw}=6.3179\times 10^{-3}\,{e^2\over r_s^3},
\end{equation}
as reported in Ref.\onlinecite{Harris}.

For the actual IBM electron density of a finite slab of thickness
$d=x\,\lambda_F/2$, introduction of Eq. (\ref{eqden}) into Eq. (\ref{xlda})
yields the ratio $\sigma_x^{LDA}/\left[\sigma_x^{LDA}\right]_{iw}$ shown in Fig.
5, as a function of $x$. As in the case of the kinetic surface energy, we
observe a considerable quantum-size effect, even for slab widths as large as
five times the Fermi wavelength. The amplitude of the oscillations is also found
to decay linearly with the thickness of the slab. However, explicit expressions
for the enveloping lines $\left[\sigma_x^{LDA}\right]_1$ and
$\left[\sigma_x^{LDA}\right]_2$ are not available; hence, for each integer value
of $x$ ($x=n$) we take
$\left[\sigma_x^{LDA}\right]_1=\sigma_x^{LDA}(n)$ and approximate
$\left[\sigma_x^{LDA}\right]_2$ as follows
\begin{equation}
\left[\sigma_x^{LDA}\right]_2\sim{\left[\sigma_x^{LDA}\right]_2^-
+\left[\sigma_x^{LDA}\right]_2^+\over 2},
\end{equation}
where $\left[\sigma_x^{LDA}\right]_2^-=\sigma_x^{LDA}(n-1/2)$ and
$\left[\sigma_x^{LDA}\right]_2^+=\sigma_x^{LDA}(n+1/2)$. Assuming that the
infinite-width limit can be extrapolated as in Eq. (\ref{sigmalim}), we use the
relation
\begin{equation}\label{xlimit2}
\left[\sigma\right]_{iw}\sim{\left[\sigma\right]_2^-+
\left[\sigma\right]_1+\left[\sigma\right]_2^+\over 3},
\end{equation}
with $\sigma$ replaced by $\sigma_x^{LDA}$.

This procedure yields the results represented in Fig. 5 by open circles,
which quickly approach the actual infinite-width limit of Eq. (\ref{xlimit1}).
The normalized LDA exchange surface energies obtained from Eq. (\ref{xlimit2})
are also shown in Fig. 6 (open circles), but now for values of $x$ up to $100$
and together with local maxima and minima,
$\left[\sigma_x^{LDA}\right]_1/\left[\sigma_x^{LDA}\right]_{iw}$ and
$\left[\sigma_x^{LDA}\right]_2^\pm/\left[\sigma_x^{LDA}\right]_{iw}$,
respectively (dashed lines). While the amplitude of the quantum-size effect
amounts, for slab widths as large as 50 times the Fermi wavelength ($x=100$), to
$1\%$ of the actual infinite-width limit, for slab widths of only 5 times the
Fermi wavelength ($x=10$) the numerical error introduced by the use of Eq.
(\ref{xlimit2}) is found to be within
$0.02\%$.

Within the IBM, we have also calculated {\it exact} exchange surface
energies by introducing Eqs. (\ref{uniform}) and (\ref{eq1}) into
Eq. (\ref{eqxc}) with $\chi^0(q,\omega)$ and
$\chi^0(z,z';q_\parallel,\omega)$ instead of
$\chi^\lambda(q,\omega)$ and $\chi^\lambda(z,z';q_\parallel,\omega)$. We
have found an oscillatory behavior of this {\it exact} contribution to the
surface energy that is similar to that shown in Fig. 5, and have tested the
approximation of Eq. (\ref{xlimit2}) (with
$\sigma$ replaced by $\sigma_x$) for slab widths as large as $100$ times
the Fermi wavelength ($x=200$). We have found
\begin{equation}\label{xexact}
\left[\sigma_x\right]_{iw}=4.0700\times 10^{-3}\,{e^2\over r_s^3},
\end{equation}
which is in exact agreement with the result obtained by Harris and
Jones\cite{Harris} for the semi-infinite medium.

The extrapolated {\it exact} exchange surface energies, as obtained from Eq.
(\ref{xlimit2}) with $\sigma$ replaced by $\sigma_x$, are represented in Fig.
6 versus $x$ (only integer values of $x$ have been considered) by open squares,
showing that it approaches the infinite-width limit of Eq. (\ref{xexact}).
Although the extrapolation procedure dictated by Eq. (\ref{xlimit2}) does not
converge as rapidly for $\sigma_x$ as for $\sigma_x^{LDA}$, the numerical error
introduced by the use of this procedure to evaluate $\sigma_x$ is still found to
be within
$0.7\%$ for slab widths of only 5 times the Fermi wavelength ($x=10$).

\subsection{Exchange-correlation surface energy}

For a given electron density, the LDA xc surface energy is obtained by
introducing into Eq. (\ref{eqxclda}) the xc energy per particle of a uniform
electron gas, $\varepsilon_{xc}^{unif}$, which we have calculated with use
of either the Wigner interpolation formula,\cite{Wigner} the Perdew-Zunger
parametrization\cite{pz} of the QMC calculation of Ceperley and
Alder,\cite{Ceperley} or Eq. (\ref{uniform}) with the RPA interacting
density-response function. We have found quantum-size
effects and the extrapolation of finite-slab calculations to the infinite-width
limit to be similar to those exhibited by kinetic and exchange contributions to
the surface energy. Hence, for each value of $r_s$ and a given number $n$ of
occupied subbands, we have considered three different values of the parameter
$x$: $n-1/2$, $n$, and $n+1/2$, and have used Eq. (\ref{xlimit2}) with $\sigma$
replaced by $\sigma_{xc}^{LDA}$. When either the Wigner interpolation formula or
the RPA are used for $\varepsilon_{xc}^{unif}$, the extrapolation procedure
dictated by Eq. (\ref{xlimit2}) yields LDA xc surface energies that are in
agreement with those reported for a semi-infinite medium in
Refs.\onlinecite{Lang75} and \onlinecite{Langreth75}, respectively. Since the
Wigner correlation energy changes very little with the electron density, as
compared with RPA or Perdew-Zunger correlation energies, it yields LDA xc surface
energies that are too small.

For each $r_s$ and given values of $x$, we have also computed {\it non-local} xc
surface energies from Eq. (\ref{eqxc}), as obtained with the xc energy of Eqs.
(\ref{uniform}) and (\ref{eq1}) computed in the RPA. We have found that the
extrapolation of these finite-slab calculations to the infinite-width limit
is accurately described by Eq. (\ref{xlimit2}), but now with $\sigma$ replaced by
$\sigma_{xc}$. Table I shows our extrapolated RPA correlation-only and
exchange-correlation surface energies, $\sigma_c$ and $\sigma_{xc}$,
together with the RPA-based LDA surface energies, $\sigma_{c}^{LDA}$ and
$\sigma_{xc}^{LDA}$, described above.\cite{procedure,notei} While the
local-density approximation overestimates, within the IBM, the {\it exact}
exchange surface energy by $55\%$ [see Eqs. (\ref{xlimit1}) and (\ref{xexact})]
for all values of
$r_s$, it largely underestimates the RPA correlation surface energy by $78\%$ at
$r_s=2$ and
$64\%$ at $r_s=6$. Large and opposite separate corrections to the LDA for
exchange and correlation nearly compensate, as discussed in
Refs.\onlinecite{Lang75} and \onlinecite{Langreth75}, and yield LDA xc surface
energies that are, over the entire metallic density range, within about $6.5\%$
of their {\it non-local} counterparts.

For comparison, Table I also shows the extrapolated LDA correlation-only and
exchange-correlation surface energies, $\sigma_{c}^{LDA'}$ and
$\sigma_{xc}^{LDA'}$, that we have obtained with use of the Perdew-Zunger
parametrization for $\varepsilon_{xc}^{unif}$. Although RPA and Perdew-Zunger
correlation energies of a uniform electron gas differ, for $r_s\sim 0$-$10$, by
$\sim 0.6$-$0.3\,{\rm eV}$ [the second-order exchange energy not present in the
RPA is known to yield $\sim 0.6\,{\rm eV}$ in the high-density limit ($r_s\to
0$)], the quantity
$\left\{\varepsilon_{xc}^{unif}[n(z)]-\varepsilon_{xc}^{unif}(\bar n)\right\}$
entering Eq. (\ref{eqxclda}) depends little on whether the RPA or the
Perdew-Zunger parametrization is used, thereby yielding LDA xc surface energies
that are comparable.   

\section{Self-consistent calculations}

Due to the simple nature of the IBM one-electron wave functions, we have used
this model as a proving ground for the calculation of both {\it local} (LDA) and
{\it non-local} xc surface energies, and also for devising the extrapolation
procedure dictated by Eq. (\ref{xlimit2}). Nevertheless, the IBM yields
unrealistic xc surface energies that are too small, especially at the lowest
values of $r_s$ that we have explored, since at these high densities the
infinite barrier constrains the electrons too close to the jellium background.

In this section, we apply Eq. (\ref{xlimit2}) in the framework of a
self-consistent microscopic model of the jellium slab. While in the IBM new
subbands for the $z$ motion become occupied every time $d$ is a multiple of
$\lambda_F/2$ ($x$ integer), for an arbitrary effective one-electron potential
new subbands are expected to enter as $2 a/\lambda_F$ (not necessarily integer)
is increased by approximately one. Accordingly, surface energies are still
oscillatory functions of the slab thickness $a$, the period of the oscillations
being approximately equal to
$\lambda_F/2$ and the amplitude decaying approximately linearly with $a$.

\subsection{LDA orbitals}

Self-consistent calculations of jellium surface energies of a semi-infinite
medium were carried out by Lang and Kohn\cite{Lang0} with use of the LDA for
exchange and correlation. They replaced the actual $v_{xc}[n](z)$ entering Eq.
(\ref{effective}) by the LDA xc potential and computed the LDA xc surface energy
from Eq. (\ref{eqxclda}) using the Wigner form\cite{Wigner} for the correlation
energy per particle of the uniform electron gas, $\varepsilon_{xc}^{unif}$. We
have tested the approximation of Eq. (\ref{xlimit2}) (with $\sigma$ replaced by
our self-consistent $\sigma_{xc}^{LDA}$), for slab widths as large as $\sim 10$
times the Fermi wavelength, and for $r_s=2$ we have found
$\left[\sigma_{xc}^{LDA}\right]_{iw}=3256\,{\rm erg/cm^2}$, in agreement with the
calculation reported by Lang and Kohn.\cite{notel} Our self-consistent
calculations of $\sigma_{xc}^{LDA}/\left[\sigma_{xc}^{LDA}\right]_{iw}$ are
plotted in Fig. 7, together with the enveloping lines
$\left[\sigma_{xc}^{LDA}\right]_1/\left[\sigma_{xc}^{LDA}\right]_{iw}$ and
$\left[\sigma_{xc}^{LDA}\right]_2/\left[\sigma_{xc}^{LDA}\right]_{iw}$, as a
function of the slab width (in units of $\lambda_F/2$). Also represented in this
figure are the extrapolated LDA xc surface energies (open circles), as obtained
from Eq. (\ref{xlimit2}) for those threshold slab widths $a_n$ for which the
$n$th subband is first occupied. One clearly sees that self-consistent density
profiles yield exactly the same oscillating behavior as obtained within the
IBM, but now QSE are smaller and the extrapolation formula of Eq.
(\ref{xlimit2}) yields a result that for a slab width as little as
$\sim 3.5$ times the Fermi wavelength ($n=7$) is within $0.01\%$ of the actual
infinite-width calculation.\cite{total}

Now we focus on our RPA-based {\it local} and {\it non-local} calculations
of the surface energy. As in the calculations of Lang and Kohn, we replace the
actual $v_{xc}[n](z)$ entering Eq. (\ref{effective}) by the LDA xc
potential,\cite{noteLDA} but with the Perdew-Zunger parametrization for the xc
potential of the uniform electron gas. {\it Local} and {\it non-local}
calculations of the xc surface energy are then carried out within one and the
same density-response framework, i.e., by either introducing Eq. (\ref{uniform})
into Eq. (\ref{eqxclda}) or by introducing both Eqs. (\ref{uniform}) and
(\ref{eq1}) into Eq. (\ref{eqxc}), with the use of the interacting RPA
density-response function. As this framework is devoid of any ambiguities in the
treatment of the many-body problem, it clearly serves as a measure of the actual
error introduced by the LDA.

In Table II, we show the infinite-width limit of the
various contributions to our RPA-based {\it local} (LDA) and {\it non-local}
jellium surface energies, as obtained from Eq. (\ref{xlimit2}) for slabs with
$n=12$ ($a\sim 5-6\,\lambda_F$).\cite{dif} As in the case of
IBM orbitals, self-consistent calculations show that the LDA
systematically overestimates/underestimates the exchange/correlation surface
energies, while the {\it local} xc surface energies differ little from their
{\it nonlocal} counterparts. Table II shows that the percentage error between
{\it local} and {\it non-local} exchange surface energies monotonically
decreases, within the present model, from a minimum of $16\%$ for the slow
varying density at
$r_s=2$ to a maximum of $95\%$ for the more rapidly varying density at $r_s=6$.
The percentage error between our {\it local} and {\it non-local} xc surface
energies is found to vary from $1.8\%$ at $r_s=2$ to $3.4\%$ at $r_s=6$.

Self-consistent electron densities vary slowly as compared to the IBM, especially
in the case of small values of $r_s$, as shown in Fig. 8. As a consequence, we
find: (a) quantum-size effects are smaller [for slabs with 12 occupied levels
the amplitude of the oscillations in both $\sigma_{xc}^{LDA}$ and $\sigma_{xc}$
ranges from $\sim 1\%$ at $r_s=2$ to $\sim 5\%$ at $r_s=6$]; (b) all
contributions to the surface energy are larger; and (c) the LDA is significantly
better, over the entire metallic range of densities ($r_s\sim 2-6$), than for
the IBM electron-density profile.

The impact on the surface energy of the short-range correlations built into the
many-body kernel of Eq. (\ref{kernel}), not present in the RPA, was investigated
in Ref.\onlinecite{Pitarke} within the ALDA. While introduction of the ALDA
kernel slightly decreases the LDA xc surface energy [the ALDA-based LDA surface
energies agree closely with the LDA quantum Monte Carlo calculations of Acioli
and Ceperley\cite{Acioli}], ALDA short-range corrections to the {\it non-local}
RPA surface energy are found to be {\it positive}. As a result, the presence of
the xc kernel enhances, within the ALDA, the discrepancy between {\it local} and
{\it non-local} xc surface energies, but the error introduced by the LDA is
still, within ALDA, of the order of $50\%$ smaller than the error one
would impute to the local approximation on the basis of the non-local FHNCA
results,\cite{Kohn86} particularly in the high-density region ($r_s=2$).

\subsection{Hartree and exchange-only orbitals}

RPA-based {\it local} and {\it non-local} surface energies, as obtained with
the use of Hartree orbitals, i.e., with the xc potential $v_{xc}[n]$ entering
Eq. (\ref{effective}) set equal to zero, were reported in
Ref.\onlinecite{Pitarke}. Hartree orbitals are more delocalized than the more
realistic LDA orbitals (with
$v_{xc}[n](z)=0$ the work function is far too small and even negative at low
values of $r_s$) and they yield a too slow varying electron-density
profile, as shown in Fig. 8. This leads to surface energies that are too
large, relative to those obtained with the use of LDA orbitals. We also
find that the impact of xc on the one-electron states is
much more pronounced than the impact of the short-range correlations built into
the many-body kernel of Eq. (\ref{kernel}).

For comparison, we have also computed kinetic, electrostatic and exchange surface
energies with an LDA exchange-only effective one-electron potential
$v_{x}[n](z)$, and have obtained the infinite-width extrapolated results shown in
Table III. A comparison of the numbers presented in Tables II and III shows that
the sum of kinetic, electrostatic and exchange surface energies is nearly
insensitive to whether an LDA exchange-only or xc effective
potential is used in the evaluation of the one-electron Kohn-Sham orbitals.
Moreover, this sum (labeled $\sigma_{HF}$ in Table III) is expected to be very
close to the {\it exact} Hartree-Fock surface energies.\cite{notehf} Our
exchange-only surface energy is for
$r_s=2.07$ in excellent agreement with the Hartree-Fock surface energy of
$-1273\,{\rm erg/cm^2}$ reported by Krotscheck and Kohn.\cite{Kohn86} Also, our
exchange-only surface energies fall for $r_s>3$ below the variational upper bound
reported by Sahni and Ma.\cite{Sahni}

\section{Summary and conclusions}

We have presented self-consistent-field calculations of the surface energy of a
bounded electron gas. First of all, we have considered the infinite-barrier
model of the surface. Within this model we have discussed the origin
of quantum-size effects, and have devised a formula that accurately extrapolates
our finite-slab calculations to the infinite-width limit. This formula may be
useful to the case of more complex slab calculations involving the use of the
band structure and/or quantum Monte Carlo simulations.

Subsequently, we have reported full self-consistent {\it non-local} calculations
of the surface energy, by employing either LDA or the less
realistic Hartree orbitals to construct the RPA density-response function. The
unambiguous nature of the comparison of {\it local} versus {\it non-local}
surface energies made possible by our self-consistent calculations leads us to
the conclusion that the local-density approximation is accurate, within
the RPA treatment of correlation. 

Our {\it non-local} RPA surface energies, as computed with the use of realistic
LDA orbitals, are in reasonable agreement with those obtained by Zhang,
Langreth, and Perdew\cite{Zhang} using the non-local Langreth-Mehl xc
functional\cite{Mehl} and also with those obtained within the recently developed
meta-generalized-gradient approximation (meta-GGA),\cite{Perdew1} which makes
use not only of the local density and its gradient but also of the orbital
kinetic-energy density. Moreover, a GGA\cite{gga} for the short-range
correlation not present in the RPA has yielded a small and {\it negative}
contribution to the surface energy,\cite{Perdew2} thereby suggesting that the
RPA may be a much better approximation for the changes in correlation energy
upon surface formation than for the total correlation energy. A recent
wave-vector interpolation as a long-range correction to the GGA,\cite{Perdew3}
which has predicted surface energies that are smaller but close to our {\it
non-local} RPA results, has indicated that the LDA
should be accurate even within an exact treatment of electron correlation. This
is in contrast with {\it non-local} correlation surface energies reported in
Refs.\onlinecite{Kohn86} and \onlinecite{Acioli}, which are significantly higher
than our {\it non-local} RPA correlation energies and the local-density
approximation.

\acknowledgments

J.M.P. acknowledges partial support by the University of the Basque Country,
the Basque Unibertsitate eta Ikerketa Saila, and the Spanish Ministerio de
Educaci\'on y Cultura. A.G.E. acknowledges support from the National Science
Foundation. The authors also acknowledge I. Sarria for his help in the
preparation of the figures. ORNL is managed by Lockheed Martin Energy Research
Corp. for the U.S. DOE under Contract No. DE-AC05-96OR22464.

\appendix

\section{}

Here we give an explicit expression for the xc surface energy of a jellium slab,
in terms of the eigenvalues $\varepsilon_l$ and the Fourier coefficients of the
eigenfunctions $\phi_l(z)$. It is also demonstrated that there is no
contribution to the xc surface energy coming from large values of the momentum
transfer parallel to the surface, so the xc energy per particle at point
$z$ is devoid of divergences.

We first introduce the following double-cosine Fourier representations for the
density-response function and the Dirac $\delta$ function:
\begin{equation}
\chi(z,z';q_\parallel,\omega)=\sum_{m=0}^\infty\sum_{n=0}^\infty
\chi_{mn}(q_\parallel,\omega)\,\cos(m\pi\tilde z)\,\cos(n\pi\tilde z')
\end{equation}
and
\begin{equation}
\delta(z-z')=\sum_{n=0}^\infty{\mu_m\over d}\cos(n\pi\tilde z)\,\cos(n\pi\tilde
z'),
\end{equation}
where $d=a+2z_0$, $\tilde z=z/d+1/2$, and $\mu_m$ is defined by the equation
\begin{equation}
\mu_m=\cases{1&for $m=0$,\cr\cr
2&for $m\ge 1$.}
\end{equation}

The electron density can also be expressed in a cosine representation
\begin{equation}
n(z)=\sum_{m=0}^\infty n_m\,\cos(m\pi\tilde z).
\end{equation}
Eq. (\ref{eqden}) yields
\begin{equation}
n_m={\mu_m\over\pi d}\sum_{l=1}^{l_M}(E_F-\varepsilon_l)G_{ll}^m,
\end{equation}
where
\begin{equation}
G_{ll'}^m=d\int_0^dd\tilde z\cos(m\pi\tilde z)\,\phi_l(\tilde
z)\,\phi_{l'}(\tilde z),
\end{equation}
or, in terms of the Fourier coefficients $b_{ls}$ of $\phi_l(z)$,  
\begin{equation}
G_{ll'}^m={1\over 2}\sum_{s=1}^\infty\sum_{s'=1}^\infty b_{ls}b_{l's'}
\left(\delta_{m,s-s'}+\delta_{m,s'-s}-\delta_{m,s+s'}\right).
\end{equation}

Then, substituting Eq. (\ref{eq1}) into Eq. (\ref{eqxc}), we obtain the
following expression for the xc surface energy:
\begin{equation}
\sigma_{xc}={1\over 2}\int{d{\bf q}_\parallel\over (2\pi)^2}
\sum_{m=0}^\infty\sum_{n=0}^\infty\alpha_{mn}(q_\parallel)
-\bar n\,a\,\varepsilon_{xc}(\bar n),
\end{equation}
where
\begin{eqnarray}\label{alpha}
\alpha_{mn}(q_\parallel)=&&v_{mn}(q_\parallel)\left[-{\hbar\over\pi}
\int_0^1d\lambda\int_0^\infty
d\omega\,\chi_{mn}^\lambda(q_\parallel,i\,\omega)\right.\cr\cr
&&\left.-{\mu_m\mu_n\over\pi
d^2}\,\sum_{l=1}^{l_M}(E_F-\varepsilon_l)\sum_{l'=1}^\infty G_{ll'}^m\,
G_{ll'}^n\right]
\end{eqnarray}
and
\begin{eqnarray}\label{coulomb}
v_{mn}(q_\parallel)=&&{2\pi e^2\over q_\parallel^2+(m\pi/d)^2}
\left[{2d\over\sqrt{\mu_m\mu_n}}\,\delta_{mn}\right.\cr\cr
&&\left.-\left[1+(-1)^{m+n}\right]
{q_\parallel\left[1-(-1)^m{\rm e}^{-q_\parallel
d}\right]\over(q_\parallel^2+(n\pi/d)^2}\right].
\end{eqnarray}

As for the Fourier coefficients $\chi_{mn}^\lambda(q_\parallel,\omega)$, Eq.
(\ref{chi}) yields
\begin{eqnarray}\label{eqintegral}
\chi_{mn}^\lambda&&(q_\parallel,\omega)=\chi_{mn}^0(q_\parallel,\omega)+
\sum_{m',n'}\chi_{mm'}^0(q_\parallel,\omega)\cr\cr
&&\times\left\{\lambda\,v_{m'n'}(q_\parallel)+
f_{m'n'}^{xc,\lambda}[n](q_\parallel,\omega)\right\}
\chi_{n'n}^\lambda(q_\parallel,\omega),
\end{eqnarray}
where the Coulomb-interaction Fourier coefficients $v_{mn}(q_\parallel)$ are
given by Eq. (\ref{coulomb}).  An explicit expression for the coefficients
$\chi_{mn}^0(q_\parallel,\omega)$ can be found in
Ref.\onlinecite{Eguiluz}. $f_{mn}^{xc,\lambda}[n](q_\parallel,\omega)$ are the
Fourier coefficients of the xc kernel. In the RPA these coefficients are simply
zero. In the ALDA,
\begin{eqnarray}
f_{mn}^{xc,\lambda,ALDA}[n](q_\parallel,\omega)=
d\int_0^d&&d\tilde z\left.{dv_{xc}(n)\over
dn}\right|_{n=n(z)}\cr\cr
&&\times\cos(m\pi\tilde z)\cos(n\pi\tilde z).
\end{eqnarray}

If the Coulomb correlations entering Eq. (\ref{eqintegral}) are ignored
altogether ($\lambda=0$), the integrations over $\lambda$ and $\omega$ in
Eq. (\ref{alpha}) can be performed analytically to find, after some algebra, the
following result:
\begin{eqnarray}
\alpha_{mn}(q_\parallel)=&&v_{mn}(q_\parallel)\,{\mu_m\mu_n\over\pi
d^2}\,\sum_{l=1}^{l_M}\sum_{l'=1}^\infty\cr\cr
&&\times\left[F_{ll'}(q_\parallel)-(E_F-\varepsilon_l)
\right]G_{ll'}^m G_{ll'}^n,
\end{eqnarray}
where
\begin{eqnarray}
F_{ll'}&&(q_{\parallel})=\varepsilon_l\,{\rm
sgn}[a_{ll'}(q_\parallel)]+\Theta[b_{ll'}(q_\parallel)]\cr\cr
&&\times\left[{a_{ll'}(q_\parallel)\sqrt{b_{ll'}(q_\parallel)}\over\pi
q_\parallel^2}
-{2\over\pi}\,\varepsilon_l\,\arctan{\sqrt{b_{ll'}(q_\parallel)}\over
a_{ll'}}\right],
\end{eqnarray}
\begin{equation}
a_{ll'}(q_\parallel)={\hbar\over 2
m}\,q_\parallel^2-{1\over\hbar}(\varepsilon_l-\varepsilon_{l'}),
\end{equation}
and
\begin{equation}
b_{ll'}(q_\parallel)=2\,q_\parallel^2\,\varepsilon_l-a_{ll'}^2.
\end{equation}

In the limit as $q_\parallel\to\infty$, we find
\begin{equation}
\lim_{q_\parallel\to\infty}F_{ll'}(q_\parallel)=(E_F-\varepsilon_l).
\end{equation}
As in this limit the interacting density-response function
approaches its non-interacting counterpart, this result shows, explicitly, that
there is no contribution to the xc surface energy coming from high values of the
parallel component of the momentum transfer, and that the singular term
$\delta(z-z')$ in Eq. (\ref{eq1}) gives no divergences.

\narrowtext

\begin{table}
\caption{Non-local RPA correlation ($\sigma_c$) and exchange-correlation
($\sigma_{xc}$) surface energies obtained in the infinite-barrier model of the
surface, and their local counterparts. Also shown are the LDA xc surface energies
($\sigma_{xc}^{LDA'}$) obtained by using the Perdew-Zunger interpolation formula
for the correlation energy of a uniform electron gas. Units are ${\rm
erg/cm^2}$. } 
\begin{tabular}{lcccccc}
%&\multicolumn{3}{c}{correlation}&\multicolumn{3}{c}{exchange-correlation}\\
$r_s$&$\sigma_{c}$&$\sigma_{c}^{LDA}$&$\sigma_c^{LDA'}$
&$\sigma_{xc}$&$\sigma_{xc}^{LDA}$&$\sigma_{xc}^{LDA'}$\\
\tableline
2.0&679&146&127&1471&1376&1357\\
2.07&617&136&118&1331&1245&1227\\
3.0&219&60&52&454&424&416\\
4.0&99&31&26&198&185&180\\
5.0&53&18&15&104&97&94\\
6.0&33&12&10&62&58&56\\
\end{tabular}
\label{table1}
\end{table}

\begin{table}
\caption{Kinetic ($\sigma_s$), electrostatic ($\sigma_{es}$), exchange
($\sigma_x$), non-local RPA xc ($\sigma_{xc}$), and total ($\sigma$) surface
energies and their local counterparts, as obtained with the use of
Perdew-Zunger parametrized LDA orbitals. Units are ${\rm erg/cm^2}$.} 
\begin{tabular}{lcccccccc}
%&\multicolumn{3}{c}{correlation}&\multicolumn{3}{c}{exchange-correlation}\\
$r_s$&$\sigma_{s}$&$\sigma_{es}$&$\sigma_x$
&$\sigma_x^{LDA}$&$\sigma_{xc}$&$\sigma_{xc}^{LDA}$&$\sigma$
&$\sigma^{LDA}$\\
\tableline
2.00&-5495&1276&2624&3037&3467&3405&-752&-814\\
2.07&-4643&1072&2296&2674&3064&3007&-507&-564\\
2.30&-2750&627&1521&1809&2098&2054&-25&-69\\
2.66&-1316&299&854&1051&1240&1211&223&194\\
3.00&-703&164&526&669&801&781&262&242\\
3.28&-435&105&364&477&579&563&249&233\\
4.00&-139&42&157&222&278&269&181&172\\
5.00&-30&17&57&92&119&115&106&102\\
6.00&-3&9&22&43&58&56&64&62\\
\end{tabular}
\label{table1}
\end{table}

\begin{table}
\caption{As in Table II, but with use of exchange-only LDA orbitals.} 
\begin{tabular}{lcccccc}
%&\multicolumn{3}{c}{correlation}&\multicolumn{3}{c}{exchange-correlation}\\
$r_s$&$\sigma_{s}$&$\sigma_{es}$&$\sigma_x$
&$\sigma_x^{LDA}$&$\sigma_{HF}$&$\sigma_{HF}^{LDA}$\\
\tableline
2.00&-5707&1390&2726&3131&-1591&-1186\\
2.07&-4832&1172&2390&2767&-1270&-899\\
3.00&-770&189&568&707&-13&126\\
4.00&-169&49&180&243&60&123\\
5.00&-46&19&71&105&44&78\\
6.00&-13&9&32&52&28&48\\
\end{tabular}
\label{table1}
\end{table}

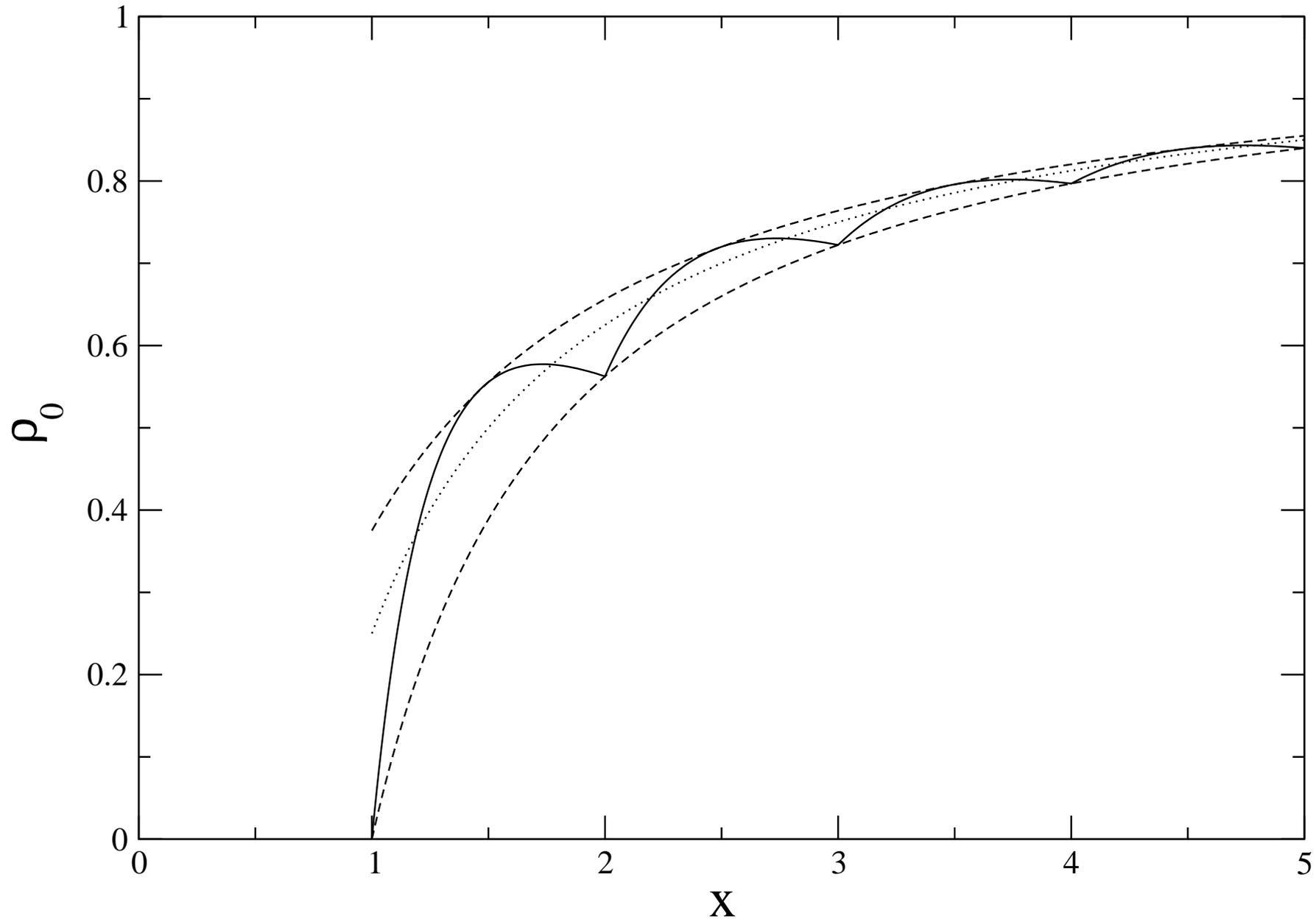
\begin{figure}
\caption{The solid line represents the normalized average electron density
$\rho_0$, as obtained from Eq. (\ref{n0}), versus the parameter
$x=2\,d/\lambda_F$. Dashed lines represent
$\left[\rho_0\right]_1$ and $\left[\rho_0\right]_2$, as obtained from Eqs.
(\ref{xn}) and (\ref{xnplus}). The dotted line represents the infinite-width
limit of Eq. (\ref{limit}).}
\end{figure}

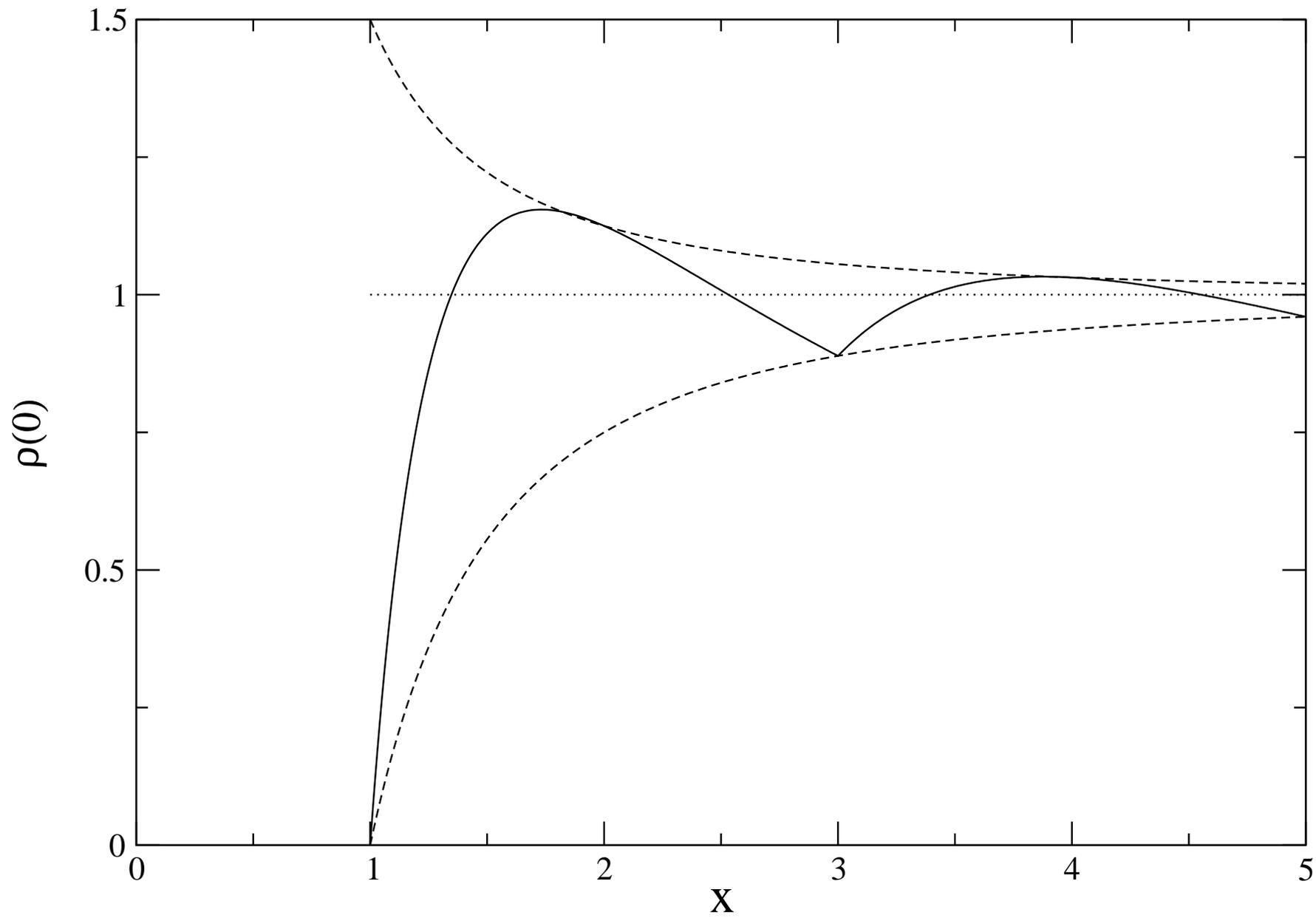
\begin{figure}
\caption{As in Fig. 1, for the normalized electron density at the center of the
slab, obtained from Eqs. (\ref{rho0}), (\ref{rho1}), and (\ref{rho2}). The
dotted line represents the infinite-width limit, $\left[\rho(0)\right]_{iw}=1$.}
\end{figure}

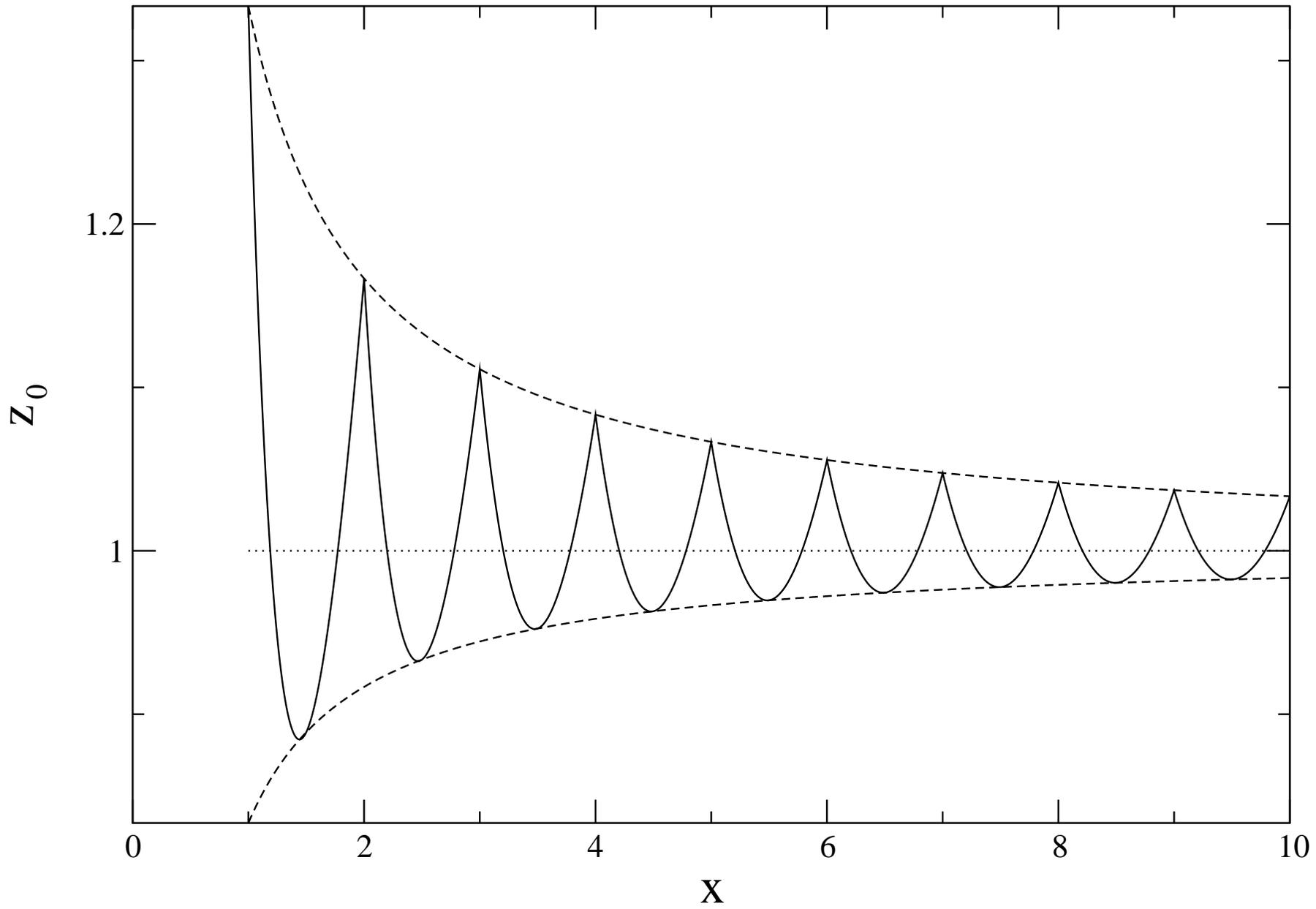
\begin{figure}
\caption{As in Fig. 1, for $z_0/\left[z_0\right]_{iw}$ obtained from Eqs.
(\ref{z0}), (\ref{z1}), and (\ref{z2}).}
\end{figure}

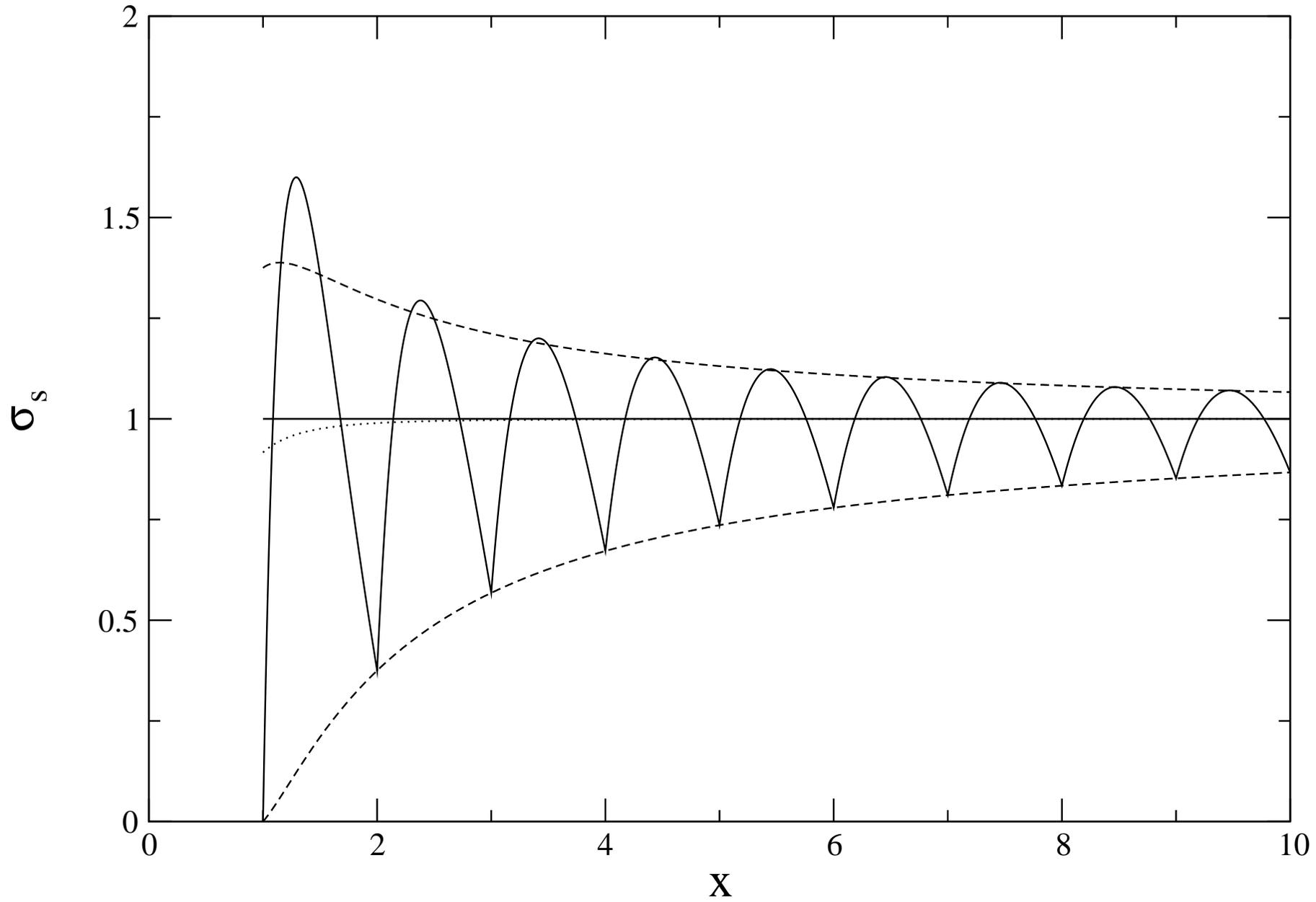
\begin{figure}
\caption{As in Fig. 1, for $\sigma_s/\left[\sigma_s\right]_{iw}$ obtained from
Eqs. (\ref{sigmas}), (\ref{env1}), and (\ref{env2}).}
\end{figure}

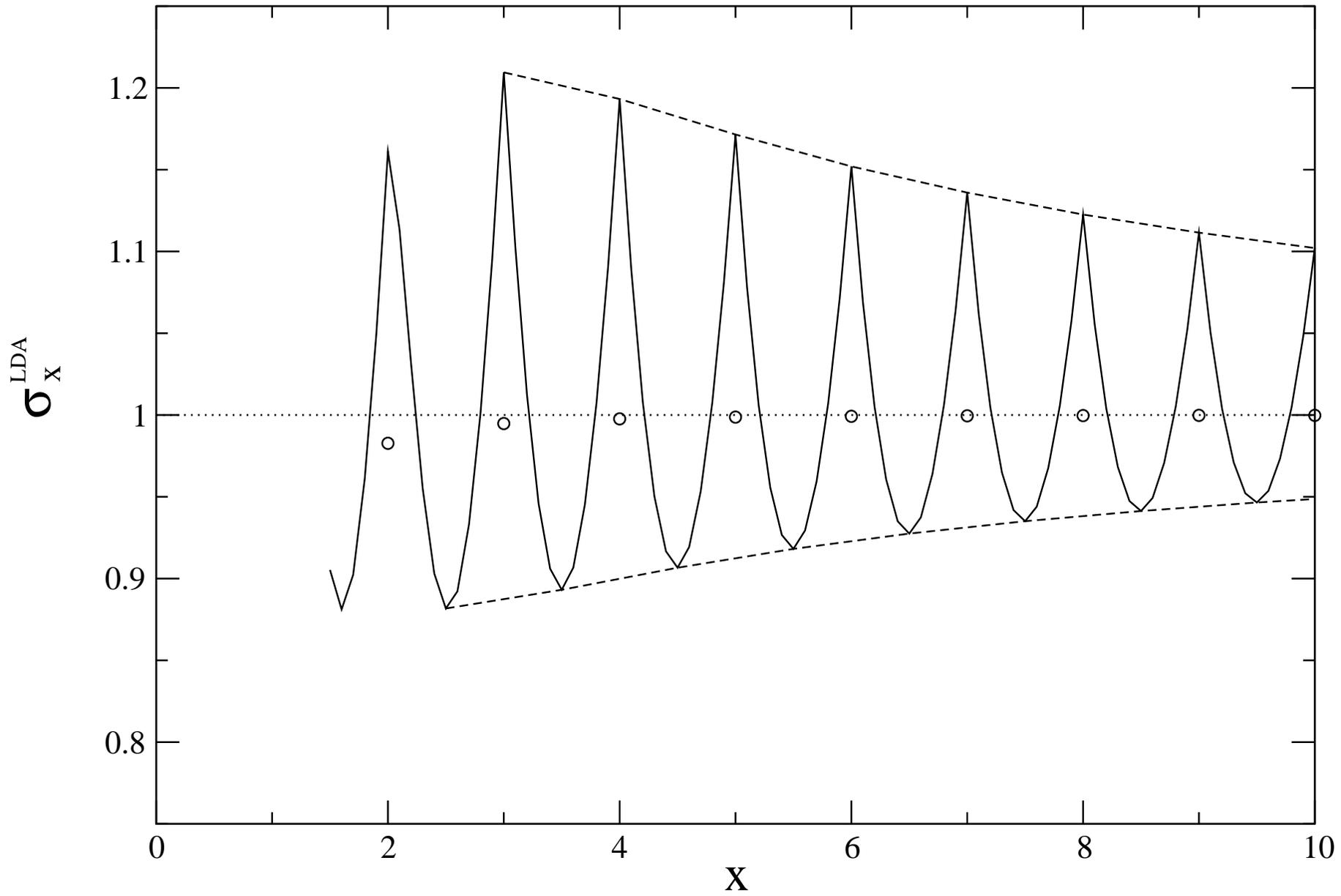
\begin{figure}
\caption{The solid line represents the normalized LDA exchange surface energy,
$\sigma_x^{LDA}/\left[\sigma_x^{LDA}\right]_{iw}$, as obtained from Eq.
(\ref{xlda}) for the IBM electron density of a finite slab, versus the parameter
$x=2\,d/\lambda_F$. Dashed lines represent the envelopes
$\left[\sigma_x^{LDA}\right]_1/\left[\sigma_x^{LDA}\right]_{iw}$ and
$\left[\sigma_x^{LDA}\right]_2^\pm/\left[\sigma_x^{LDA}\right]_{iw}$. The
extrapolated normalized LDA surface energies, as obtained from Eq.
(\ref{xlimit2}) with $\sigma$ replaced by $\sigma_x^{LDA}$, are represented by
open circles.}
\end{figure}

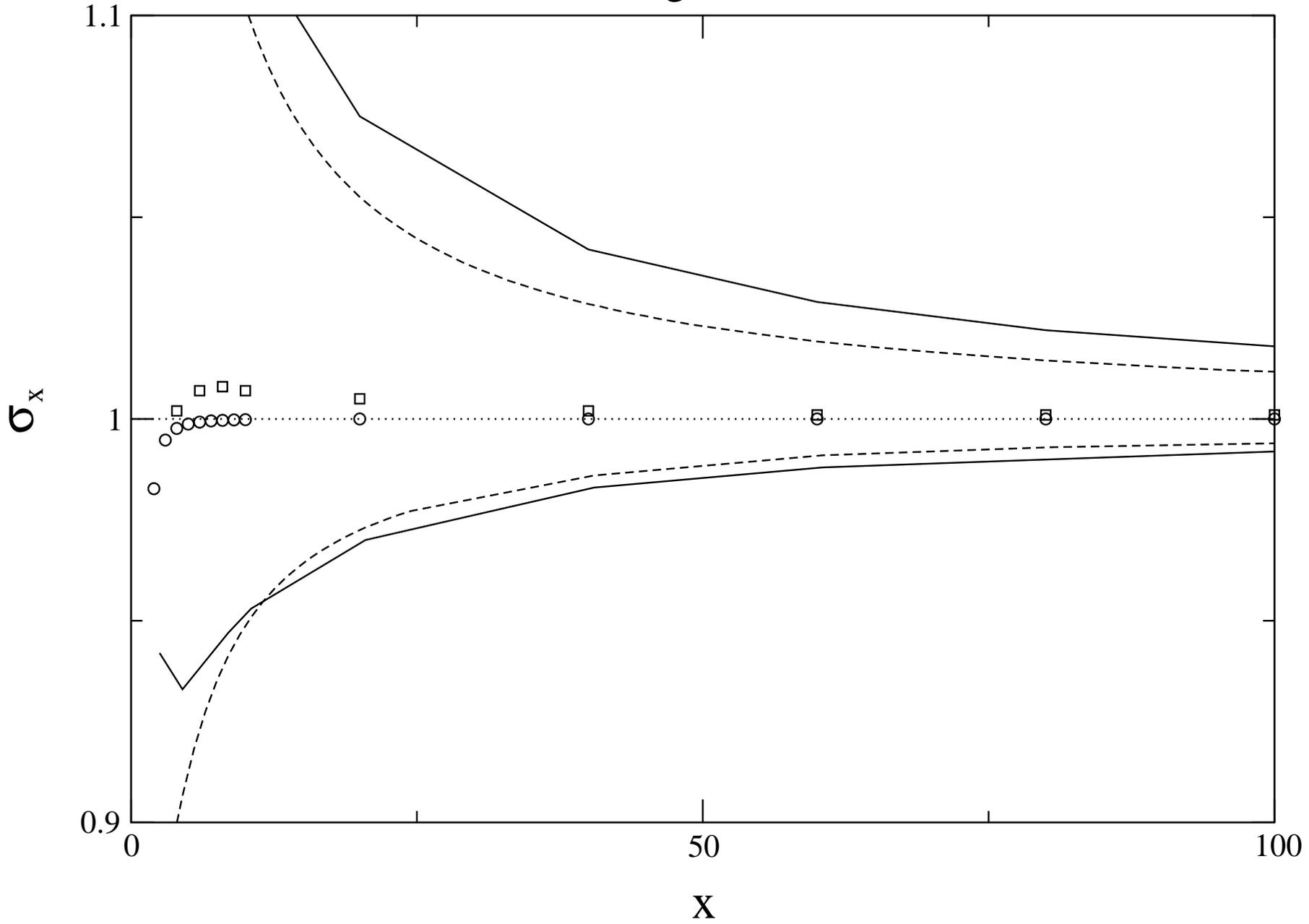
\begin{figure}
\caption{The solid lines represent the envelopes
$\left[\sigma_x\right]_1/\left[\sigma_x\right]_{iw}$ and
$\left[\sigma_x^{LDA}\right]_2^\pm/\left[\sigma_x\right]_{iw}$ of the normalized
exact exchange surface energies, which have been obtained, within
the IBM, by introducing Eqs. (\ref{uniform}) and (\ref{eq1}) into Eq.
(\ref{eqxc}) with $\chi^0(q,\omega)$ and
$\chi^0(z,z';q_\parallel,\omega)$ instead of $\chi^\lambda(q,\omega)$ and
$\chi^\lambda(z,z';q_\parallel,\omega)$. The corresponding LDA enveloping lines
(also shown in Fig. 5) are represented by dashed lines. Open squares and circles
represent extrapolated exact (squares) and LDA (circles) exchange surface
energies, as obtained from the procedure dictated by Eq. (\ref{xlimit2}).}
\end{figure}

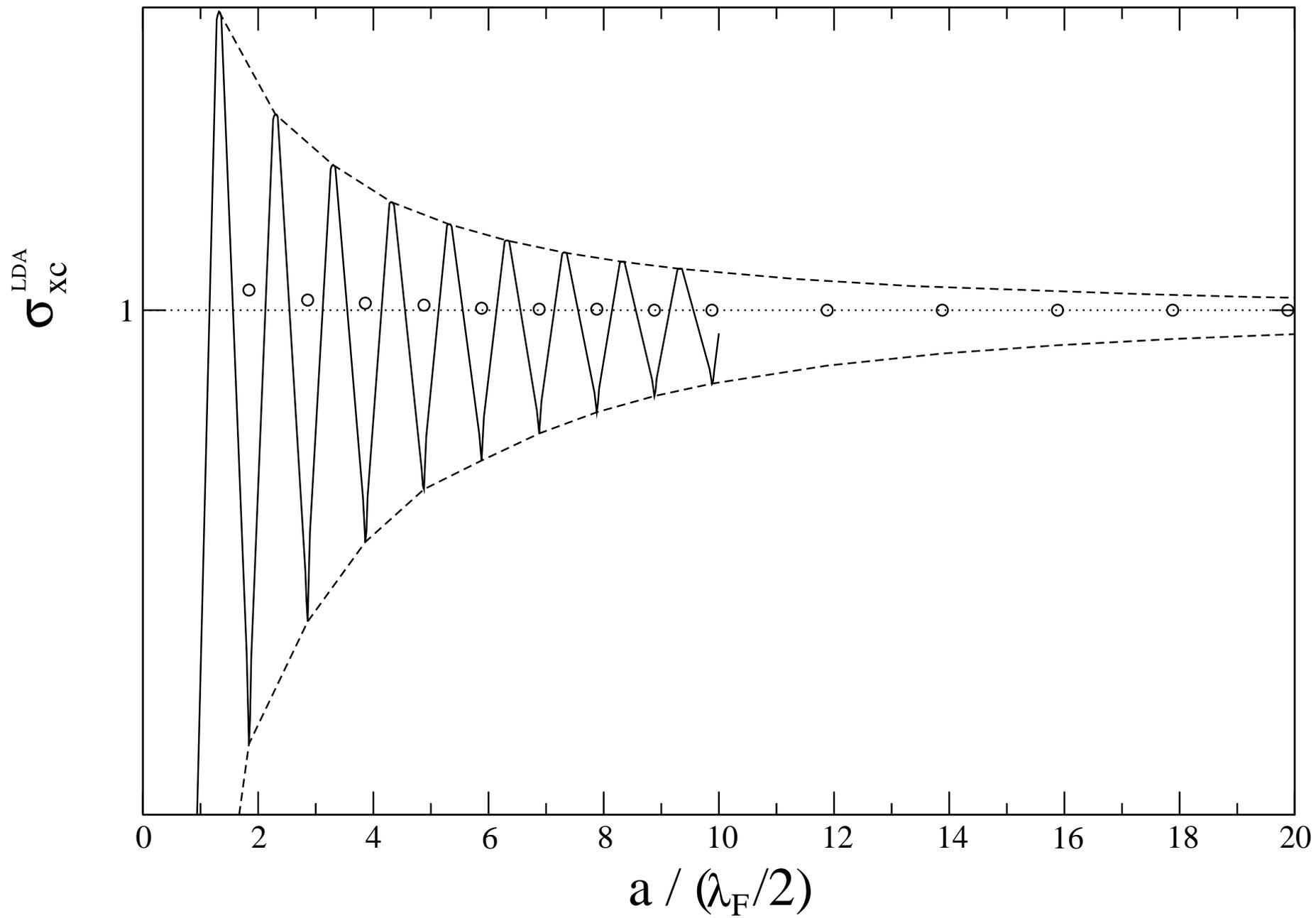
\begin{figure}
\caption{The solid line represents the normalized LDA xc surface energy,
$\sigma_{xc}^{LDA}/\left[\sigma_{xc}^{LDA}\right]_{iw}$, as obtained from Eq.
(\ref{eqxclda}) for the Lang-Kohn electron-density profile of a finite slab,
versus the slab width (in units of $\lambda_F/2$). Dashed lines represent the envelopes
$\left[\sigma_x^{LDA}\right]_1/\left[\sigma_x^{LDA}\right]_{iw}$ and
$\left[\sigma_x^{LDA}\right]_2^\pm/\left[\sigma_x^{LDA}\right]_{iw}$. The
extrapolated normalized LDA surface energies, as obtained from Eq.
(\ref{xlimit2}) with $\sigma$ replaced by $\sigma_{xc}^{LDA}$, are represented by
open circles.}
\end{figure}

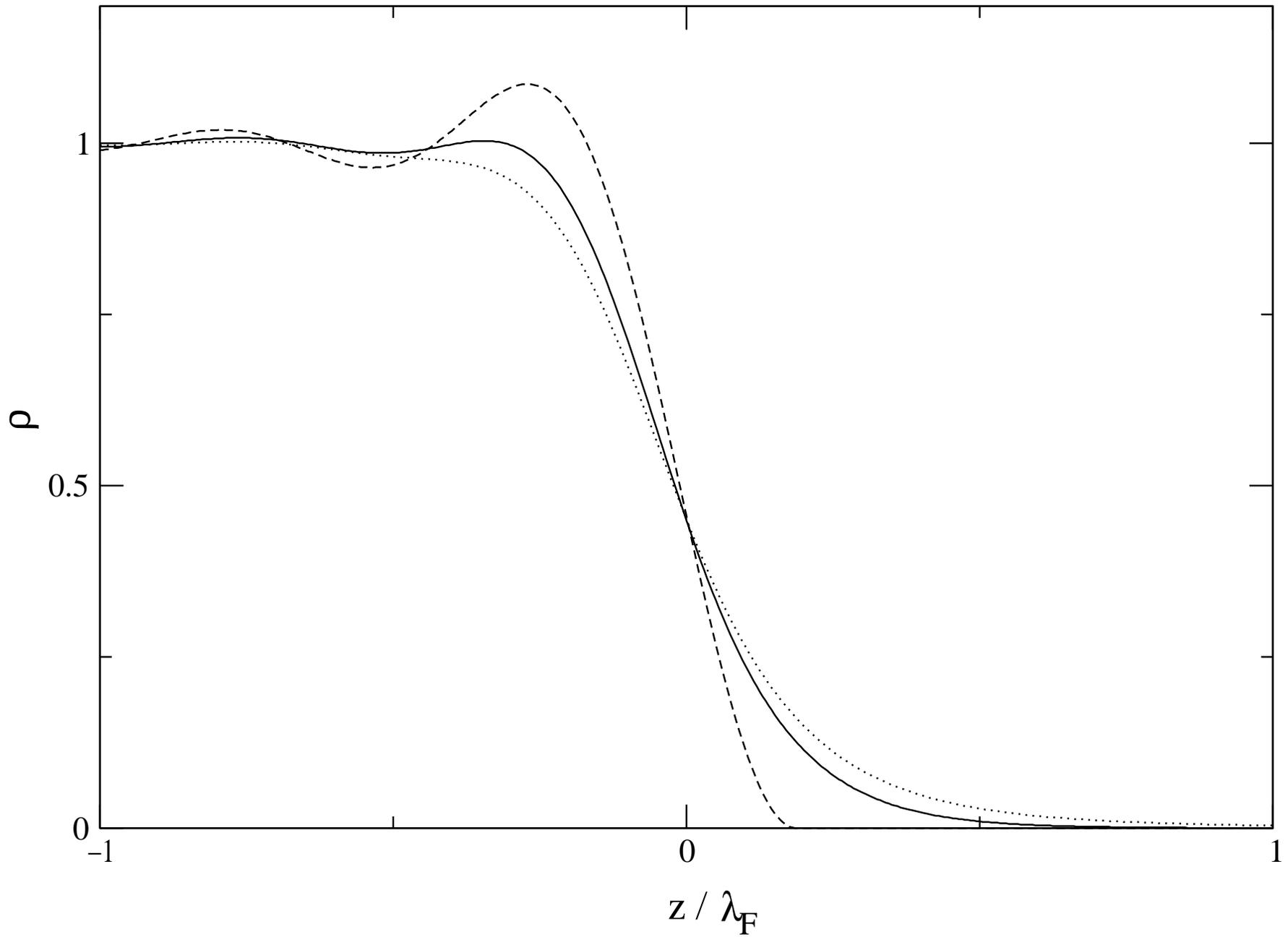
\begin{figure}
\caption{The normalized electron-density profile of a jellium surface with
$r_s=2.07$, as obtained with use of Perdew-Zunger parametrized LDA orbitals
(solid line), Hartree orbitals (dotted line), and within the IBM (dashed line).
We note that the normalized IBM electron-density profile, with the $z$
coordinate measured in units of the Fermi wavelength, is independent of $r_s$.
In this figure, the jellium edge has been taken at $z=0$.}
\end{figure}


\begin{references}
\bibitem{note1} In the non-selfconsistent infinite-barrier model (IBM), $z_0$ is
chosen so as to ensure charge neutrality. For a self-consistent description of
the density-response function, $z_0$ is chosen sufficiently large for the
physical results to be insensitive to the precise value employed.
\bibitem{noteHarris} Since the so-called electron-density parameter
$r_s$, defined by the relation $1/\bar n=(4/3)\pi r_s^3$, is usually given in
units of the Bohr radius, $a_0$, we note that $e^2/a_0^3=1.5569\times 10^6\,{\rm
erg}/{\rm cm}^2$.
\bibitem{procedure} We estimate that the numerical error introduced by our slab
calculations corresponds to one unit in the last digit of all the entries in
Tables I, II, and III.
\bibitem{notei} Small differences between our calculated xc surface energies and
those reported by Wikborg and Inglesfield\cite{Wikborg} are due to the
fact that contributions from $q_\parallel>2q_F$ to the integral of Eq.
(\ref{eq1}) were omitted by these authors.
\bibitem{total} Due to the subtle cancellations that exist among the various
contributions to the surface energy, the introduction of small errors in the
kinetic, electrostatic, and xc terms may result in larger uncertainties of the
total surface energy.
\bibitem{noteLDA} The use of the LDA xc potential and, therefore, LDA orbitals,
should not be confused with the use of the LDA to replace
$\varepsilon_{xc}[n](z)$ in Eq. (\ref{eqxc}) by the xc energy per particle of a
uniform electron gas of density $n(z)$, $\varepsilon_{xc}^{unif}[n(z)]$.
\end{references}
\end{document}